\documentclass[12pt,twoside]{article}

\usepackage{fancyheadings}
\usepackage{graphics}
\advance \headheight by 3.0truept       

\usepackage{cite}   
\usepackage{epsf}

\usepackage{times}

\newcommand{\lt}{\left}
\newcommand{\rt}{\right}
\newcommand{\nn}{\nonumber \\}

\newcommand{\eq}[1]{(\ref{#1})}
\newcommand{\bra}[1]{\ensuremath{\langle \, #1 \, |}}
\newcommand{\ket}[1]{\ensuremath{| \, #1 \, \rangle }}
\def\nicefrac#1#2{\hbox{${#1\over #2}$}}

\begin{document}
\thispagestyle{plain}
~\\[-20truemm]
FERMILAB-Pub-00/084-T, \hfill 
\parbox{3cm}{hep-ph/0004139}\\
~\vspace{0.5truecm}
\begin{center}
{\LARGE\bf 
$\bf{ B_{s,d} \rightarrow \ell^+ \ell^-}$ in a Two-Higgs-Doublet Model 
}\\[2\baselineskip]
\textsl{
Heather E. Logan\footnote{electronic addresses: logan@fnal.gov, nierste@fnal.gov }
and \addtocounter{footnote}{-1}
Ulrich Nierste\footnotemark\\ 
Theoretical Physics Department, Fermilab, Batavia, IL 60510-0500, 
USA.\footnote{Fermilab is operated by Universities Research Association Inc.\
under contract no.~DE-AC02-76CH03000 with the US Department of Energy} \\[2mm]
}
\end{center}
\vspace{5mm}
\begin{center}

\textbf{\large Abstract} 
\end{center}
We compute the branching fractions of $B_{s,d} \to \ell^+ \ell^-$ in the
type-II two-Higgs-doublet model with large $\tan\beta$.  
We find that the parameters of the neutral Higgs sector of the 
two-Higgs-doublet model cancel in the result,
so that the branching fractions depend only on the charged
Higgs mass and $\tan\beta$.
For large values of $\tan\beta$ and a charged Higgs mass above the
bound from $b \to s \gamma$, we find that the branching fractions 
can be enhanced by up to an order of magnitude or suppressed by up to a 
factor of two compared to the Standard Model result.
We point out that previous calculations in the literature are
gauge-dependent due to the omission of an important diagram, which
gives the dominant contribution in the 't Hooft-Feynman gauge.  
We have analyzed in detail the region of the $(M_{H^+}, \tan \beta)$
plane to be probed by searches for $B_s \to \mu^+ \mu^-$ in
Run II of the Tevatron. Since the branching fraction increases like
$\tan^4 \beta$, this decay mode is complementary to $ b \to s
\gamma$ and efficiently probes the large $\tan \beta$ region. For
$\tan \beta=60$, an integrated luminosity of 20 fb$^{-1}$ in an 
extended Run II will
probe charged Higgs masses up to 260 GeV, if the background to 
$B_s \to \mu^+ \mu^-$ is small. 
For the same value of $\tan\beta$, the LHC may be able to explore charged
Higgs masses up to 1 TeV using this decay.  \\[2mm] 
{\small PACS: 13.20.He, 12.60.Fr, 14.80.Cp}
\begin{center}
-------------------------------------------------------------------
\end{center}

\section{Introduction}
The ongoing and forthcoming high-statistics B-physics experiments at
BaBar, BELLE, HERA-B, the Tevatron, and the LHC experiments ATLAS,
CMS and LHCb \cite{exp} will probe the
flavor sector of the Standard Model (SM) with high precision. These
experiments may reveal physics beyond the SM, and a
substantial theoretical effort is devoted to calculating the
observables that will be tested in various scenarios of new physics. 

A common feature of
all popular weakly-coupled extensions of the SM is an
enlarged Higgs sector. In this paper we study the type-II
two-Higgs-doublet model (2HDM), which has the same particle 
content and tree-level Yukawa couplings as 
the Higgs sector of the
Minimal Supersymmetric Model (MSSM).
If the ratio $\tan\beta$ of the two Higgs vacuum expectation values 
is large, the Yukawa
coupling to $b$ quarks is of order one and large effects on $B$ decays
are possible.  Direct searches for the lightest neutral MSSM Higgs
particle have begun to constrain the low $\tan \beta$ region in the MSSM, 
because the
theoretically predicted mass range increases with $\tan \beta$. Hence
observables allowing us to study the complementary region of large $\tan
\beta$ are increasingly interesting.  A further theoretical motivation
to study the large $\tan \beta$ case is SO(10) grand unified
theories \cite{SO10}: they unify the top and bottom Yukawa couplings at high
energies, corresponding to $\tan \beta$ of order 50.

The leptonic decay $B_{d^{\prime}} \rightarrow \ell^+ \ell^-$,
where $d^{\prime} = d$ or $s$ and $\ell = e$, $\mu$ or $\tau$,
is especially well suited to the study of an enlarged Higgs sector
with large $\tan\beta$.
In the SM the decay amplitude suffers from a
helicity-suppression factor of $m_{\ell}/m_b$, which is absent in the
Higgs-mediated contribution. 
This helicity suppression factor numerically competes with the
suppression factor of $(m_{\ell}/M_W)\tan\beta $ stemming from the
Higgs Yukawa couplings to the final state leptons, so that one expects the
new contributions in the 2HDM to be similar in size to those of the SM.

Earlier papers have already addressed the decay
$B_{d^\prime}\rightarrow \ell^+ \ell^-$ in the 2HDM or the full MSSM
\cite{hnv,savage,sk,dhh,hly,cg}. Yet the presented results differ analytically
and numerically substantially from each other, so that we have decided
to perform a new analysis.

This paper is organized as follows.  In section \ref{sec:2HDM} we give
a brief review of the type-II 2HDM.  In section \ref{sec:EHcalc} we
review the SM calculation of the decay $B_{d^{\prime}} \to \ell^+ \ell^-$
and describe our calculation of the
relevant 2HDM diagrams.  We finish section \ref{sec:EHcalc} by
combining the results for the 2HDM diagrams and giving compact
expressions for the branching fractions.  In section
\ref{sec:comparisons} we compare our result with previous
calculations.  In section \ref{sec:phenomenology} we present a
numerical analysis of our result and estimate the reach of future
experiments.  We present our conclusions in section
\ref{sec:conclusions}.  Finally the appendix contains a discussion of
trilinear Higgs couplings.

\section{The two-Higgs-doublet model}
\label{sec:2HDM}
In this paper we study the type-II 2HDM.  The model is reviewed
in detail in ref. \cite{HHG}.
The 2HDM contains two complex SU(2) doublet scalar fields,
\begin{equation}
	\Phi_1 = \left(\begin{array}{c}
		\phi_1^+ \\
		\phi_1^0  \end{array} \right),
	\qquad
	\Phi_2 = \left(\begin{array}{c}
		\phi_2^+  \\
		\phi_2^0  \end{array} \right),
\end{equation}
which acquire the vacuum expectation values (vevs) 
$\langle \phi_i^0 \rangle = v_i$ and break the electroweak symmetry.
The Higgs vevs $v_1$ and $v_2$ are constrained by the $W$ boson mass, 
$M^2_W = \nicefrac{1}{2}g^2(v_1^2 + v_2^2)
= \nicefrac{1}{2}g^2v^2_{SM}$, where $v_{SM} = 174$ GeV is the SM Higgs
vev.  Their ratio is parameterized by $\tan\beta = v_2/v_1$.

Since in this paper we are not interested in CP-violating quantities, 
we assume CP is conserved by the Higgs sector for simplicity.
The mass eigenstates are then given as follows.
The charged Higgs states are
\begin{eqnarray}
	G^+ &=& \phi_1^+ \cos\beta + \phi_2^+ \sin\beta \nonumber \\
	H^+ &=& -\phi_1^+ \sin\beta + \phi_2^+ \cos\beta,
\end{eqnarray}
and their hermitian conjugates.  The CP-odd states are
\begin{eqnarray}
	G^0 &=& \phi_1^{0,i} \cos\beta + \phi_2^{0,i} \sin\beta \nonumber \\
	A^0 &=& -\phi_1^{0,i} \sin\beta + \phi_2^{0,i} \cos\beta,
\end{eqnarray}
where we use the notation $\phi^0_i = v_i + \nicefrac{1}{\sqrt{2}}
(\phi_i^{0,r} + i\phi_i^{0,i})$
for the real and imaginary parts of $\phi^0_i$.
The would-be Goldstone bosons $G^{\pm}$ and $G^0$ are eaten 
by the $W$ and $Z$ bosons.
The CP-even states mix by an angle $\alpha$ giving two states,
\begin{eqnarray}
	H^0 &=& \phi_1^{0,r} \cos\alpha + \phi_2^{0,r} \sin\alpha \nonumber \\
	h^0 &=& -\phi_1^{0,r} \sin\alpha + \phi_2^{0,r} \cos\alpha.
\end{eqnarray}

In order to avoid large flavor-changing neutral Higgs interactions
we require natural flavor conservation \cite{GlashowWeinbergPaschos}.
We impose the discrete symmetry $\Phi_1 \to -\Phi_1$,
$\Phi_2 \to \Phi_2$ (which is softly
broken by dimension-two terms in the Higgs potential), with the SU(2)
singlet fermion fields transforming as $d \to -d$, $u \to u$, $e \to -e$.
These transformation rules define the type-II 2HDM and determine the
Higgs-fermion Yukawa couplings.
The Yukawa terms in the Lagrangian are: 
\begin{equation}
	\mathcal{L}_{\rm Yuk} = -Y_d \bar Q \Phi_1 d 
	- Y_u \bar Q \Phi_2^c u - Y_l \bar L \Phi_1 e + {\rm h.c.}
\end{equation}
where $\Phi^c = i \tau_2 \Phi^*$.  Down-type
quarks and charged leptons (up-type quarks) are given mass by their 
couplings to $\Phi_1$ ($\Phi_2$).

Most of the Higgs couplings needed in our calculation are given 
in ref. \cite{HHG}.  In addition we must consider the trilinear
$H^+H^-H$ couplings ($H=h^0,H^0$) which were first given in ref.
\cite{sk} and are discussed in appendix \ref{app:trilinear}.

\section{Effective hamiltonian for $\mathbf{B \rightarrow \ell^+ \ell^-}$}
\label{sec:EHcalc}
The decay $B_{d^{\prime}} \to \ell^+ \ell^-$
proceeds through loop diagrams and is of fourth order in the weak 
coupling.  In both the SM and 2HDM, the contributions with a top quark
in the loop are dominant, so that one may describe the decay at low energies
of order $m_b$ by a local $\bar b d^{\prime} \bar \ell \ell$ coupling 
via the effective hamiltonian,
\begin{equation}
	H = \frac{G_F}{\sqrt{2}} 
	\frac{\alpha_{EM}}{2\pi \sin^2\theta_W} \xi_t
	\left[ C_S Q_S + C_P Q_P + C_A Q_A \right].
	\label{hami}
\end{equation}
Here $G_F$ is the Fermi constant, $\alpha_{EM}$ is the electromagnetic 
fine structure
constant and $\theta_W$ is the Weinberg angle. The CKM elements are
contained in $\xi_t= V_{tb}^* V_{td^{\prime}}$.  The operators in
\eq{hami} are
\begin{equation}
	Q_S \; = \; m_b \, \bar b P_L d^{\prime} \, \bar \ell \ell,  \qquad
	Q_P \; = \; m_b \, \bar b P_L d^{\prime} \, \bar \ell \gamma_5 \ell,
		\qquad  
	Q_A \; = \; \bar b \gamma^{\mu} P_L d^{\prime} \, 
		\bar \ell \gamma_{\mu} \gamma_5 \ell,
	\label{ops}
\end{equation}
where $P_L = (1-\gamma_5)/2$ is the left-handed projection operator.
We have neglected the right-handed scalar quark operators because
they give contributions only proportional to the $d^{\prime}$ mass.
The vector leptonic operator $\bar \ell \gamma^{\mu} \ell$ does not
contribute to $B_{d^\prime}\rightarrow \ell^+ \ell^-$, because it
gives zero when contracted with the $B_{d^{\prime}}$ momentum.
Finally, no operators involving 
$\sigma_{\mu \nu}=i \, [\gamma_{\mu} ,\gamma_{\nu}]/2$ contribute
to $B_{d^{\prime}} \to \ell^+ \ell ^-$. 

Because $m_b \ll M_W,m_t,M_{H^+}$, there are highly separated mass
scales in the decay $B_{d^{\prime}} \to \ell^+ \ell^-$.
Short-distance QCD corrections can therefore contain large logarithms
like $\log (m_b/M_W)$, which must be summed to all orders in
perturbation theory with the help of renormalization group techniques.
The calculation of the diagrams in the full high-energy theory gives
the initial condition for the Wilson coefficients at a high
renormalization scale $\mu$ on the order of the heavy masses in the
loops. The hadronic matrix elements, however, are calculated at a low
scale $\mu={\cal O} (m_b)$ characteristic of the $B_{d^{\prime}}$
decay.  The renormalization group evolution down to this low scale
requires the solution of the renormalization group equations of the
operators $Q_A$, $Q_S$ and $Q_P$.  Yet the operator $Q_A$ has zero
anomalous dimension because it is a ($V - A$) quark current, which is
conserved in the limit of vanishing quark masses.  Similarly, the
operators $Q_S$ and $Q_P$ have zero anomalous dimension because the
anomalous dimensions of the quark mass $m_b (\mu)$ and the (chiral)
scalar current $\bar b P_L d^\prime (\mu)$ add to zero. Hence the
renormalization group evolution is trivial: if the bottom quark mass
in $Q_S$ and $Q_P$ is normalized at a low scale $\mu={\cal O} (m_b)$,
then no large logarithms appear in the effective hamiltonian or in the decay
rate.

\begin{figure}[tbp]
	\begin{center}
	\resizebox{8cm}{8cm}
	{\includegraphics*[53mm,131mm][163mm,244mm]{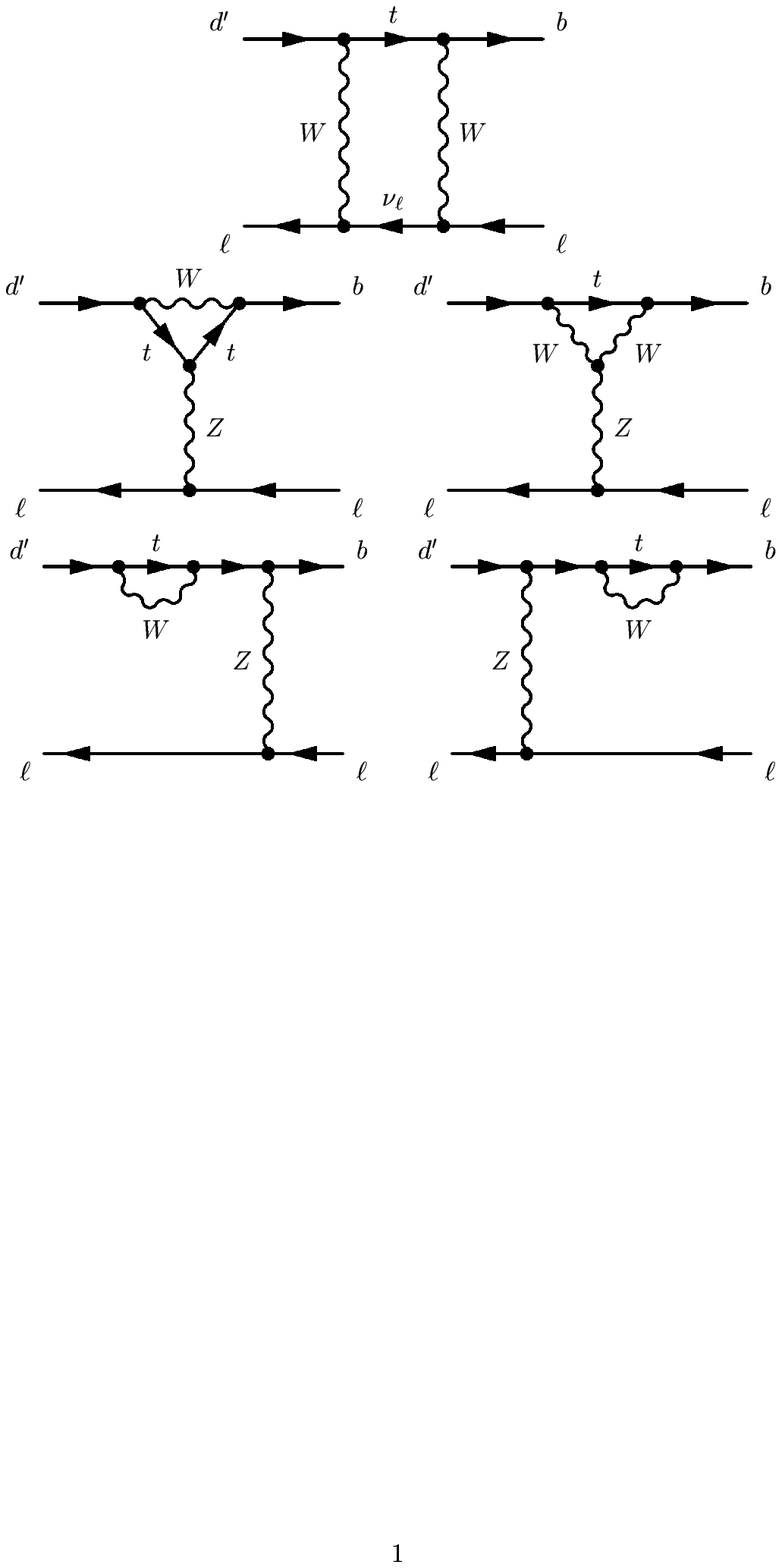}}
	\end{center}
	\caption{Dominant SM diagrams.}
	\label{fig:SMdiags}
\end{figure}

In the SM, the dominant contributions to this decay come
from the $W$ box and $Z$ penguin diagrams shown in fig.\ \ref{fig:SMdiags}.
These diagrams were first calculated
in \cite{InamiLim} and give a non-negligible contribution only to the
Wilson coefficient $C_A$.  There is no contribution from 
a photonic penguin because of the photon's pure vector coupling to leptons.
There are also contributions to the Wilson coefficient $C_S$ from a 
SM Higgs penguin \cite{GK}
and to the Wilson coefficient $C_P$ from the would-be neutral Goldstone 
boson penguin \cite{Krawczyk},
but these contributions to the amplitude are suppressed by a factor
of $m_b^2/M_W^2$ relative to the dominant contributions and can be ignored.

The SM decay amplitude is then given by the Wilson coefficient \cite{InamiLim}
\begin{equation}
	C_A = 2 Y(x_t),
	\label{eq:CA}
\end{equation}
where $x_t = m_t^2(m_t)/M_W^2 = 4.27 \pm 0.26$ and $m_t$ is evaluated in the
$\overline{\rm MS}$ scheme at $\mu = m_t$, giving $m_t(m_t) = 166$ GeV.  
The function $Y(x_t)$ is given by 
$Y(x_t) = Y_0(x_t) + \frac{\alpha_s}{4\pi}Y_1(x_t)$, where
$Y_0(x_t)$ gives the leading order (LO) contribution calculated in 
\cite{InamiLim} and $Y_1(x_t)$ incorporates the next-to-leading (NLO) 
QCD corrections and is 
given in \cite{bbmu}.  The NLO corrections increase $Y(x_t)$ by about
3\%, if $m_t$ is normalized at $\mu=m_t$.  Numerically,
\begin{equation}
	Y(x_t) = 0.997 \left[\frac{m_t(m_t)}{166 \, {\rm GeV}}\right]^{1.55},
	\label{eq:Y}
\end{equation}
where we have parameterized the dependence on the running top quark
mass in the $\overline{\rm MS}$ scheme.

\begin{figure}[tbp]
	\begin{center}
	\resizebox{9cm}{!}
	{\includegraphics*[2.03in,6.56in][6.53in,9.58in]{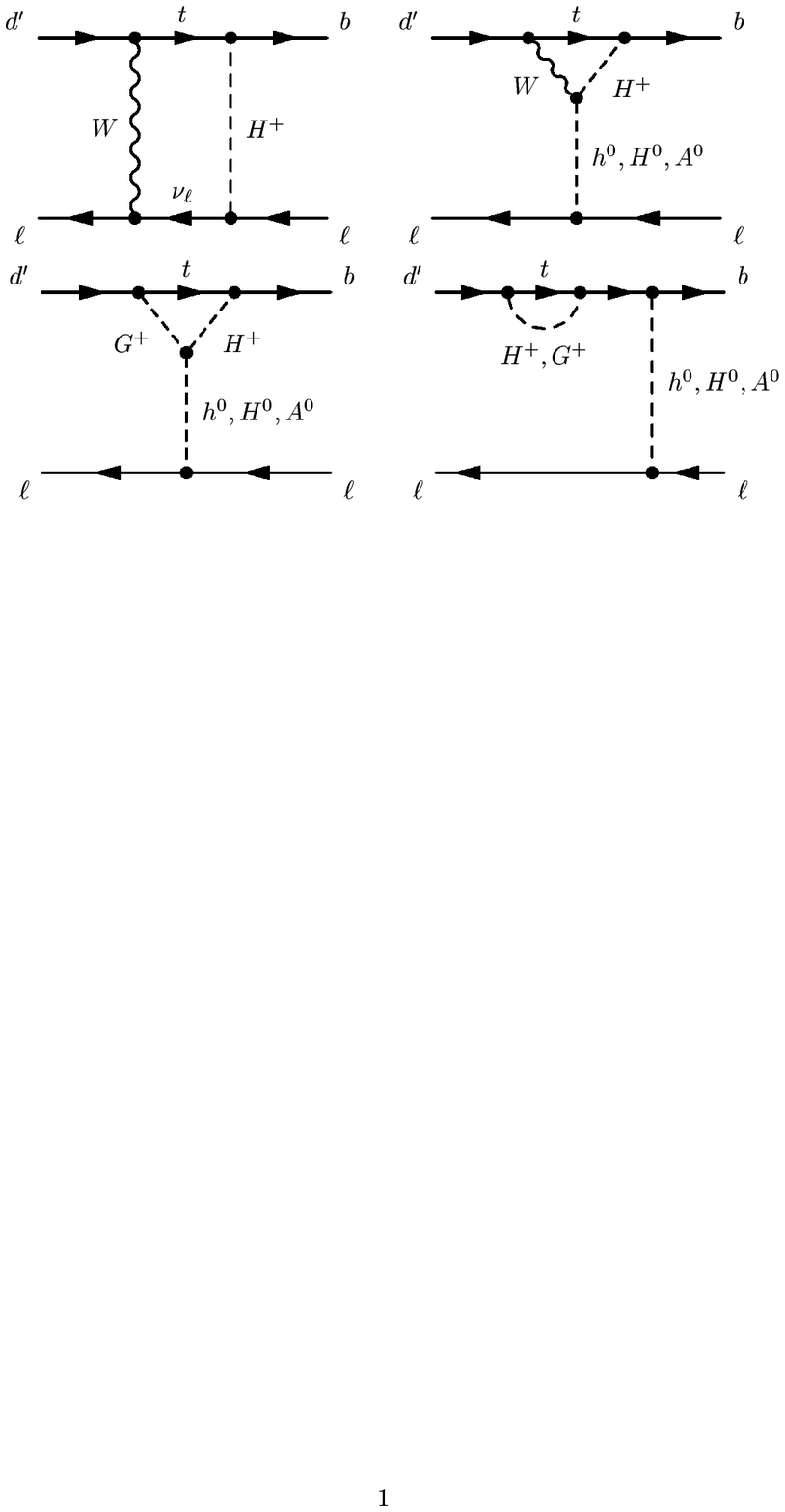}}
	\end{center}
	\caption{Dominant diagrams in the 2HDM with large $\tan\beta$.}
	\label{fig:2HDMdiags}
\end{figure}

We limit our consideration to the case of large $\tan\beta$, for 
which the 2HDM contributions to this decay are significant.
In the large $\tan\beta$ limit, the Wilson coefficients $C_P$ and $C_S$ 
receive sizeable contributions from the box diagram involving $W$ 
and $H^+$ and the penguins and fermion self-energy diagrams with 
neutral Higgs boson exchange shown in fig.\ 
\ref{fig:2HDMdiags}.
There are no new contributions to $C_A$ in the 2HDM, which therefore
retains its SM value.

We have calculated the individual diagrams explicitly in a general
$R_{\xi}$ gauge, keeping only the terms proportional to 
$\tan^2\beta$.  Although each diagram that involves a $W^{\pm}$
or $G^{\pm}$ boson is gauge-dependent, their sum is 
gauge-independent.  For compactness, we give the results of the 
individual diagrams below in the 't~Hooft-Feynman gauge.

\subsection{Box diagram}
The box diagram in fig.~\ref{fig:2HDMdiags} gives the following
contribution to $C_S$ and $C_P$:
\begin{eqnarray}
	C_S^{box} &=& C_P^{box} \;= \; 
       \frac{m_{\ell}}{2 M_W^2} \tan^2 \beta \, B_+ \! \lt( x_{H^+}, x_t \rt). 
	\label{box}
\end{eqnarray}
Here $x_{H^+} = M^2_{H^+}/M^2_W$ and $x_t$ was defined after
\eq{eq:CA} in terms of $m_t(m_t)$.  Strictly speaking, in a LO
calculation like ours, one is not sensitive to the renormalization
scheme and we could equally well use the top quark pole mass in $x_t$.
However, experience with NLO calculations in the SM \cite{bbmu}
shows that with the definition of $m_t$ adopted here, higher-order
QCD corrections are small in leptonic decays. Finally the loop
function $B_+$ in \eq{box} reads
\begin{eqnarray}
	B_+ \! \lt( x , y \rt) &=& \frac{y}{x-y}
		 \lt[ \frac{\log y}{y-1} - \frac{\log x}{x-1} \rt].
	\label{boxf}
\end{eqnarray}
$B_+ ( x_{H^+}, x_t ) $ also contains the contribution from internal
up and charm quarks with $m_c=m_u=0$ from the implementation of the 
GIM mechanism.  The effect of a nonzero charm quark mass is negligibly
small.

\subsection{Penguins}
\label{sect:penguins}
The penguin diagram with $H^+$ and $W^+$ in the loop (see
fig.~\ref{fig:2HDMdiags}) contributes
\begin{eqnarray}
	C_S^{peng,1} &=& 
    	\frac{m_{\ell}}{2} \, \tan^2 \beta \, P_+ \! \lt( x_{H^+}, x_t \rt) 
    	\lt[ \frac{\sin^2 \alpha}{M_{h^0}^2} + 
         \frac{\cos^2 \alpha}{M_{H^0}^2}	\rt], \nn
	C_P^{peng,1} &=& 
    	\frac{m_{\ell}}{2} \, \tan^2 \beta \, P_+ \! \lt( x_{H^+}, x_t \rt) 
    	\frac{1}{M_{A_0}^2} . 
	\label{peng1} 
\end{eqnarray}
Here again all three quark flavors enter the result from the GIM
mechanism, and the effect of nonzero charm quark mass is negligible.
By contrast, in the penguin diagram involving $H^+$ and $G^+$ in the
loop only the internal top quark contribution is relevant, because the
coupling of $G^+$ to quarks is proportional to either of the quark
masses and we neglect $m_s$.  This diagram gives
\begin{eqnarray}
	C_S^{peng,2} &=& 
	-\frac{m_{\ell}}{2} \, \tan^2 \beta \, 
    	P_+ \! \lt( x_{H^+}, x_t \rt) 
  	\lt[ \, \frac{\sin^2 \alpha}{M^2_{h^0}} 
	\frac{(M_{H^+}^2 - M_{h_0}^2)}{M_W^2}  \rt. \nn
	&& \lt. 
 \phantom{ - \frac{m_{\ell}}{2 M_W^2} \tan^2 \beta P_+ \lt( x_{H^+}, x_t \rt)} 
        + \frac{\cos^2 \alpha}{M^2_{H^0}}
	\frac{(M_{H^+}^2 - M_{H_0}^2)}{M_W^2}
	\rt], \nn
	C_P^{peng,2} &=& 
   	-\frac{m_{\ell}}{2} \, \tan^2 \beta \, 
     	P_+ \! \lt( x_{H^+}, x_t \rt) \frac{1}{M^2_{A^0}}
	\lt[ \, \frac{M_{H^+}^2 - M_{A^0}^2}{M_W^2} \rt] . 
	\label{peng2} 
\end{eqnarray}
The results in \eq{peng1} and \eq{peng2} involve the loop function
\begin{eqnarray}
 	P_+ \! \lt( x , y \rt) &=& \frac{y}{x-y}
	\lt[ \frac{x \, \log x}{x-1} - \frac{y\, \log y}{y-1} \rt] .
 	\label{pengf}
\end{eqnarray}

\subsection{Self-energies}
Before we write down the result from the diagrams with self-energies
in the external quark lines, we discuss how these contributions come
into play.  A treatment of flavor-changing self-energies has been
proposed in \cite{ahkkm,ms} and analyzed in some detail in \cite{ds}.
In these works flavor-changing self-energies have been discussed in
a different context, the renormalization of the $W$-boson coupling to
quarks.  In \cite{ahkkm,ms,ds} counterterms have been chosen in such a
way that the flavor-changing self-energies vanish if either of the
involved quarks is on-shell.  
For the down-type
quarks these counterterms form two $3\times 3$ matrices in flavor
space, one for the left-handed quark fields and one for the
right-handed ones, and similarly for the up-type quarks.
It was argued in \cite{ms,ds} that the
anti-hermitian parts of the counterterm matrices for the left-handed
fields can be absorbed into a renormalization of the CKM matrix, and
the hermitian parts of the matrices can be interpreted as wave
function renormalization matrices $Z^L_{ij}$ and $Z^R_{ij}$ with
$i,j=d,s,b$ for our case of external down-type quarks.

However, it has also been argued \cite{ggm} that the on-shell scheme
of \cite{ms,ds} is not gauge invariant.  In addition we find that the
approach of \cite{ahkkm,ms,ds} leads to an inconsistency in our
calculation.  We cannot cancel the anti-hermitian parts of the
self-energies in the external lines with the counterterms for the CKM
matrix, because unlike in the case of the $W$ coupling renormalization
there is no tree-level coupling of a neutral scalar or vector boson
to $\bar b d^{\prime}$ to be renormalized.  Hence we cannot absorb the
anti-hermitian parts of the flavor-changing self-energy matrices
into counterterms and they do contribute to our calculation. 

The absorption of the hermitian parts into wave function counterterm
matrices is optional, because the introduction of wave function
counterterms only trivially shuffles self-energy contributions into
vertex counterterms.  It is most straightforward then to avoid the
issue of counterterms altogether by simply calculating the fermion
self-energy diagrams as they are shown in fig.\ \ref{fig:2HDMdiags}.
This calculation is straightforward because the internal $b$ quark
line is off-shell and therefore it does not contribute to the
1-particle pole of the $s$ quark and needs not be truncated. It is
crucial to note that one must start with $m_s \neq m_b$ in the
diagrams with external self-energies in fig.~\ref{fig:2HDMdiags} to
properly account for the quark propagator $1/(m_b-m_s)$, and then take the
limit $m_b,m_s \rightarrow 0$ (except in the $\tan \beta$-enhanced
Higgs couplings, of course) at the end. FCNC transitions become
meaningless for degenerate quark masses, and one obtains an incorrect result
if one starts with $m_s = m_b$ and regulates the propagator pole with
an off-shell momentum $p$ with $p\rightarrow 0$. In this respect the
Higgs exchange diagrams in fig.~\ref{fig:2HDMdiags} differ from the
situation with $\gamma$- or $Z$- exchanges, where both methods give
the same correct result. Further we note that for $m_s \neq m_b$
one must include flavor-changing
self-energies in external lines with a factor of 1 rather than of 1/2
as in the flavor-conserving case. This is due to the fact that
flavor-conserving self-energies come from the residue of the
one-particle pole, while in our approach flavor-changing self-energies
are part of the non-truncated Green's function. By close inspection of
the formulae in \cite{ds} one also recovers this ``factor of 1 rule''
from the expressions for the wave function renormalization matrices
derived in \cite{ds}. 

There are two fermion self-energy diagrams that contribute in the
2HDM, one with a would-be Goldstone boson $G^+$ and one with the
physical charged Higgs $H^+$. Their sum is ultraviolet-finite. This is
different from the SM case, in which the UV divergence of the
$G^+$ diagram cancels with the UV divergence of a SM Higgs vertex diagram
involving a $G^+$ and a top quark in the loop.  
As in the penguin diagrams involving $H^+$ and $G^+$ in the loop,
only the internal top quark contributions to the self-energy are
relevant here, because the coupling of $H^+$ or $G^+$ to quarks is proportional
to either of the quark masses and we neglect $m_s$.
The self-energy diagrams add the following term to the Wilson coefficients:
\begin{eqnarray}
	C_S^{self} &=& 
      \frac{m_{\ell}}{2} \, \tan^2 \beta \, 
	\left( x_{H^+} - 1 \right) \, P_+ (x_{H^+}, x_t)
    	\lt[ \frac{\sin^2 \alpha}{M_{h^0}^2} + 
         \frac{\cos^2 \alpha}{M_{H^0}^2}	
	\rt], \nn
	C_P^{self} &=& 
    	\frac{m_{\ell}}{2} \, \tan^2 \beta \, 
	\left( x_{H^+} - 1 \right) \, P_+ (x_{H^+}, x_t)
    	\frac{1}{M_{A_0}^2} . 
	\label{self} 
\end{eqnarray}

\subsection{2HDM Wilson coefficients and branching ratios}
Adding \eq{box}, \eq{peng1}, \eq{peng2} and \eq{self} we obtain the
2HDM Wilson coefficients in \eq{hami}:
\begin{equation}
	C_S = C_P = \frac{m_{\ell}}{2 M_W^2} \tan^2 \beta \,\, 
	\frac{\log r}{r-1},
	\label{wc}
\end{equation}
where $r \equiv x_{H^+}/x_t = M_{H^+}^2/m_t^2(m_t)$.  Note that the
dependence on the masses of the neutral Higgs bosons from the penguin
and fermion self-energy diagrams has dropped out in their sum without
invoking any relation between the mixing angle 
$\alpha$ and the Higgs masses.  The
result depends on only two of the 2HDM parameters:
$\tan\beta$ and $M_{H^+}$.

The two hadronic matrix elements involved are related by the field
equation of motion
\begin{eqnarray}
	\bra{0} \bar b \gamma^{\mu} \gamma_5 d^{\prime} (x)
	\ket{ B_{d^{\prime}} (P_{B_{d^{\prime}}})}
	 &=& 
  	i \, f_{B_{d^{\prime}}} \, P_{B_{d^{\prime}}}^\mu 
	e^{-i \, P_{B_{d^{\prime}}} \cdot x} \nn 
	\bra{0} \bar b \gamma_5 d^{\prime} (x)
	\ket{ B_{d^{\prime}} (P_{B_{d^{\prime}}})}
	&=& 
    	- i \, f_{B_{d^{\prime}}} \, 
	\frac{m_{B_{d^{\prime}}}^2}{m_b+m_{d^{\prime}}} 
	e^{-i \, P_{B_{d^{\prime}}} \cdot x}.
\end{eqnarray}

The resulting decay amplitude is
\begin{equation}
	\left| \mathcal{A} \right| = \frac{G_F}{\sqrt{2}} 
	\frac{\alpha_{EM} m_{B_{d^{\prime}}} f_{B_{d^{\prime}}}}
	{4\pi \sin^2\theta_W} 
	\left| \xi_t
	\left[ \left( m_{B_{d^{\prime}}} C_S \right) \bar \ell \ell
	+ \left( m_{B_{d^{\prime}}} C_P 
	- \frac{2m_{\ell}}{m_{B_{d^{\prime}}}} C_A \right) 
	\bar \ell \gamma_5 \ell \right] \right| .
\end{equation}
Here $f_{B_{d^{\prime}}}$ is the $B_{d^{\prime}}$ decay constant, 
normalized according to $f_{\pi} = 132$ MeV.
The corresponding branching ratio is
\begin{eqnarray}
	\mathcal{B}(B_{d^{\prime}} \to \ell^+\ell^-) 
	&=& \frac{G_F^2 \alpha_{EM}^2}{32 \pi^2 \sin^4 \theta_W}
	\frac{m_{B_{d^{\prime}}}^3 \tau_{B_{d^{\prime}}} f^2_{B_{d^{\prime}}}}
	{8 \pi} |\xi_t|^2
	\sqrt{1 - \frac{4m_{\ell}^2}{m^2_{B_{d^{\prime}}}}}
	\nonumber \\
	&& \times
	\left[ \left(1 - \frac{4m_{\ell}^2}{m^2_{B_{d^{\prime}}}}\right)
	m^2_{B_{d^{\prime}}} C_S^2
	+ \left( m_{B_{d^{\prime}}} C_P 
	- \frac{2m_{\ell}}{m_{B_{d^{\prime}}}} C_A \right)^2 \right],	
	\label{eq:BR}
\end{eqnarray}
where $\tau_{B_{d^{\prime}}}$ is the $B_{d^{\prime}}$ lifetime.

Numerically, the branching fractions are given by
\begin{eqnarray}
	\mathcal{B}(B_d \to \ell^+ \ell^-) &=& 
	(3.0 \times 10^{-7})
	\left[ \frac{\tau_{B_d}}{1.54 \, {\rm ps}} \right]
	\left[ \frac{f_{B_d}}{210 \, {\rm MeV}} \right]^2
	\left[ \frac{|V_{td}|}{0.008} \right]^2 
	\frac{m_{\ell}^2}{m_{B_d}^2}
	\sqrt{1- \frac{4m_{\ell}^2}{m_{B_d}^2}} \nonumber \\
	&& \hspace{-2.5cm} \times 
	\left[ \left(1 - \frac{4 m_{\ell}^2}{m_{B_d}^2}\right)
	\left( \frac{m_{B_d}^2 \tan^2\beta}{8 M_W^2} 
	\frac{\log r}{r-1} \right)^2
	+ \left( \frac{m_{B_d}^2 \tan^2\beta}{8 M_W^2} 
	\frac{\log r}{r-1} - Y(x_t) \right)^2 \right], \nonumber \\
	\mathcal{B}(B_s \to \ell^+ \ell^-) &=&
	(1.1 \times 10^{-5})
	\left[ \frac{\tau_{B_s}}{1.54 \, {\rm ps}} \right]
	\left[ \frac{f_{B_s}}{245 \, {\rm MeV}} \right]^2
	\left[ \frac{|V_{ts}|}{0.040} \right]^2 
	\frac{m_{\ell}^2}{m_{B_s}^2}
	\sqrt{1- \frac{4m_{\ell}^2}{m_{B_s}^2}} \nonumber \\ 
	&& \hspace{-2.5cm} \times
	\left[ \left(1 - \frac{4 m_{\ell}^2}{m_{B_s}^2}\right)
	\left( \frac{m_{B_s}^2 \tan^2\beta}{8 M_W^2} 
	\frac{\log r}{r-1} \right)^2
	+ \left( \frac{m_{B_s}^2 \tan^2\beta}{8 M_W^2} 
	\frac{\log r}{r-1} - Y(x_t) \right)^2 \right].
	\label{eq:numBRs}
\end{eqnarray}

It is a well known property of the 2HDM that there exists a limit
in which the particles $A^0$, $H^0$, and $H^+$ become very heavy
and decouple from processes at the electroweak energy scale while
$h^0$ remains light and its couplings approach those of the SM
Higgs particle \cite{decoupling}.
In the limit of large $M_{H^+}$, $C_P$ and $C_S$ fall as $M_{H^+}^{-2}$.
Thus the deviation of the branching fractions from their SM prediction falls as
$M_{H^+}^{-2}$ in the large $M_{H^+}$ limit, and the effects of the 
enlarged Higgs sector decouple.

Next we discuss the accuracy of the large $\tan \beta$ approximation.
Subleading terms in $\tan \beta$ could be enhanced by powers of
$m_t/m_b$ compared to our result in \eq{wc}, as is the case for the
SM contribution. Such terms indeed occur, but they are
suppressed by two powers of $\cot \beta$ compared to the SM
terms in \eq{wc}.  Hence the formulae above are sufficient for all
purposes; e.g.\ if $\tan\beta =50$ the terms subleading in $\tan\beta$
give a correction only of $\mathcal{O} (2\%)$.
If $\tan\beta $ is between a few and 15 the 2HDM corrections are 
small and experimentally hard to resolve, so that an error of order
$\cot \beta$ is tolerable as well. 

\section{Comparison with other calculations}
\label{sec:comparisons}

\subsection{The analyses of He et al. and of Savage}
In the paper of He, Nguyen and Volkas \cite{hnv}, the decays
$B \to \ell^+ \ell^-$, $B \to K \ell^+ \ell^-$ and $b \to s \ell^+ \ell^-$
are analyzed in both type-I and type-II 2HDMs.  
In \cite{hnv}, 
the only diagrams considered are the box diagram
involving two charged Higgs bosons and the $A^0$ penguin involving
$H^+$ and $W^+$ in the loop.  Although the calculations of \cite{hnv}
are performed in the 't~Hooft-Feynman gauge, the $A^0$ penguin
involving $H^+$ and $G^+$ in the loop is not considered.
Similarly, in the paper of Savage \cite{savage}, the decay $B \to \mu^+ \mu^-$
is considered in the 2HDM, with and without tree-level flavor-changing
neutral Higgs couplings.  Only the contribution of the $A^0$ 
penguin is considered.
In both \cite{hnv} and \cite{savage}, several diagrams that are 
important at large $\tan\beta$
and required in order to obtain a gauge-independent result are neglected.

\subsection{The analysis of Skiba and Kalinowski}
In the paper of Skiba and Kalinowski \cite{sk},
the decay $B_s \to \tau^+ \tau^-$ is analyzed in the type-II 2HDM.  
In \cite{sk} additional penguin diagrams are considered that are not
proportional to $\tan^2\beta$, but rather contain one or no powers of
$\tan\beta$.  We have neglected these contributions in our analysis,
because they are not relevant for the interesting case of large
$\tan\beta$.  These additional penguin diagrams can be important for
small values of $\tan\beta$ and in regions of the parameter space where
some of the Higgs quartic couplings are very large resulting in large
trilinear $H^+H^-H$ couplings ($H=h^0,H^0,A^0$).

Considering only terms proportional to $\tan^2\beta$, our results for
the individual diagrams agree with those of \cite{sk}, with two
important exceptions.  First, the authors of \cite{sk} incorrectly
conclude that the box diagram involving $H^+$ and $W^+$
is negligible and therefore neglect it.  If we neglect the
box diagram, we find that the sum of the remaining contributions
proportional to $\tan^2\beta$ is gauge-dependent. In the
't~Hooft-Feynman gauge employed in \cite{sk} the omitted diagram gives
the dominant contribution, affecting the numerical result substantially.
Second, our result for the $A^0$ penguin diagram involving
$H^+$ and $G^+$ in the loop differs from that of \cite{sk} by a sign.
Our sign is required for the gauge-independence of $C_P$.

\subsection{The analyses of Huang et al.\ and Choudhury et al.}
In the paper of Dai, Huang and Huang \cite{dhh}, the Wilson
coefficients in \eq{hami} are calculated in the type-II 2HDM at large
$\tan\beta$.  As in our calculation, only the diagrams proportional to
$\tan^2\beta$ are considered.  However, the authors of \cite{dhh}
consider only the penguin and fermion self-energy diagrams with
neutral Higgs boson exchange and neglect the box diagram with a $W$
and charged Higgs boson in the loop.  Still, after leaving out the box
diagram, our results for $C_S$ and $C_P$ in the 't~Hooft-Feynman
gauge do not agree with those of \cite{dhh}.  This is partly due to a
typographical error in \cite{dhh}, which is corrected in \cite{hly,cg}.
There are two remaining discrepancies.  First, our result for the
$A^0$ penguin diagram involving $H^+$ and $W^+$ in the loop differs
from that of \cite{dhh} by a sign.  Second, in \cite{dhh} a
contribution from the $h^0$ and $H^0$ penguin diagrams with two $H^+$
bosons in the loop is included.  This diagram is included in
\cite{dhh} because it apparently receives a factor of $\tan\beta$ from
the trilinear $H^+H^-H$ couplings ($H=h^0,H^0$).  We argue in appendix
\ref{app:trilinear} that the trilinear couplings should not be
considered $\tan\beta$ enhanced.  Therefore we conclude that the
penguin diagram with two $H^+$ bosons in the loop should not be
included in the $\mathcal{O}(\tan^2\beta)$ calculation because it
is of subleading order in $\tan\beta$.

In \cite{hly,cg} the Wilson coefficients in \eq{hami} are calculated
for supersymmetric models with large $\tan\beta$.  If the diagrams
involving supersymmetric particles are neglected, this calculation
reduces to that for the type-II 2HDM with parameters constrained by
supersymmetric relations.  Again, in \cite{hly,cg} only the diagrams
with neutral Higgs boson exchange are considered, and the box diagram
with a charged Higgs boson and $W$ boson is not included.  Leaving out
the box diagram, our result for $C_S$ and $C_P$ in the
't~Hooft-Feynman gauge does not agree with the non-SUSY part of that
of \cite{hly,cg}.  This discrepancy arises because our result for the
penguin diagrams involving $H^+$ and $W^+$ in the loop differs from
that of \cite{hly,cg} by a sign.  Once SUSY relations are imposed on
the Higgs sector, it is clear that the penguin diagrams with two $H^+$
bosons in the loop are not of order $\tan^2\beta$, and the authors of
\cite{hly,cg} have omitted these diagrams, as we did.

A final critical remark concerns the treatment of the renormalization
group in the paper by Choudhury and Gaur \cite{cg}. They include an
additional renormalization group factor to account for the running of
the Wilson coefficients.  Yet these authors have overlooked that the
running of the (chiral) scalar quark current in $Q_S$ and $Q_P$ (see
\eq{ops}) is compensated by the running of the $b$-quark mass multiplying
the currents as explained in sect.~\ref{sec:EHcalc}. This leads to an
underestimate of the Wilson coefficients by roughly 23\%.

In conclusion, the papers in \cite{hnv,savage}, \cite{sk} and \cite{dhh,hly,cg}
disagree with each other, and our calculation does not agree
with any of them. None of the results in \cite{hnv,savage,sk,dhh,hly,cg} passes
the test of gauge-independence and, in our opinion, each contains mistakes.

\section{Phenomenology}
\label{sec:phenomenology}
As can be seen from the numerical coefficients in \eq{eq:numBRs},
$\mathcal{B}(B_s \to \ell^+\ell^-)$ is significantly larger than
the corresponding branching fraction for $B_d$.  This is due 
primarily to the relative sizes of $|V_{ts}|$ and $|V_{td}|$.
As a result, even though the production rate of $B_s$ is three
times smaller than that of $B_d$ at the Tevatron, the bounds on
the leptonic branching fractions of $B_s$ are much closer to the 
SM predictions than those of $B_d$ \cite{cdf}.  For this reason
we concentrate on the decays of $B_s$.
Because of the suppression of the branching fractions by $m_{\ell}^2/m_B^2$, 
clearly the decay to $\tau$ pairs is the largest of the leptonic branching
fractions in both the SM and the 2HDM.  
However, this decay is very difficult to reconstruct 
experimentally (due to the two missing neutrinos), 
and as a result the experimental limits on $B$ decays to $\tau$ pairs
are very weak.  Therefore in our numerical analysis we focus on the 
decay $B_s \to \mu^+ \mu^-$,
for which the experimental bound is the closest to the SM prediction.
The best experimental bound comes from CDF \cite{cdf}, where
one candidate event for $B \to \mu^+ \mu^-$ has been reported;
this event was consistent with the expected background and lay in
the overlapping part of the search windows for $B_d$ and $B_s$.
The corresponding 95\% confidence level upper bound on the 
$B_s \to \mu^+ \mu^-$ branching fraction is \cite{cdf}
\begin{equation}
	\mathcal{B}(B_s \to \mu^+\mu^-) < 2.6 \times 10^{-6} 
	\qquad ({\rm expt}).
\end{equation}
The SM prediction for the branching fraction is 
\begin{equation}
	\mathcal{B}(B_s \to \mu^+\mu^-) = 4.1 \times 10^{-9} \qquad ({\rm SM}),
\end{equation}
where we have taken the central values for all inputs in \eq{eq:numBRs} and
ignored the 2HDM contributions as well as the errors in the hadronic 
parameters.

Because the 2HDM Wilson coefficients in \eq{wc} depend on only two of
the parameters of the 2HDM, $\tan\beta$ and $M_{H^+}$, the behavior of
the result in different parts of parameter space is easy to
understand.  In regions of the parameter space with a large 2HDM contribution
to $\mathcal{B}(B_{d^{\prime}} \to \ell^+ \ell^-)$ compared
to the SM contribution, we may neglect $Y(x_t)$ in 
\eq{eq:numBRs}.  Then the result is particularly simple: the
branching fractions are proportional to $\tan^4 \beta \, \log^2 r /
(r-1)^2$.

\begin{figure}[tbp]
	\begin{center}
	\resizebox{13cm}{!}
	{\rotatebox{270}
	{\includegraphics{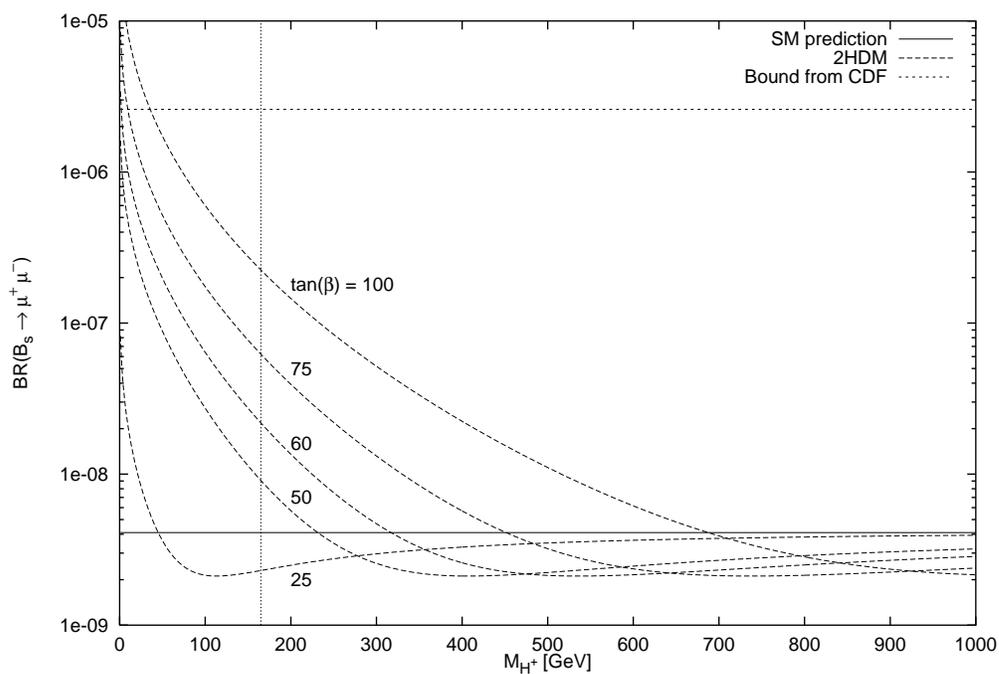}}}
	\end{center}
	\caption{$\mathcal{B}(B_s \to \mu^+\mu^-)$ in the 2HDM as a function
of $M_{H^+}$ for $\tan\beta = 100$, 75, 60, 50 and 25.  
For comparison we show the 
current experimental bound \cite{cdf} and the SM prediction for the
branching fraction.  The vertical line is the lower bound on $M_{H^+}$
in the type-II 2HDM from $b \to s \gamma$ 
\cite{btosgamma}.}
	\label{fig:BRplot}
\end{figure}

We can see from \eq{eq:numBRs} and the value of $Y(x_t)$
given in \eq{eq:Y} that the interference between the SM and 2HDM
contributions to the branching fractions is destructive.
The effect of the destructive interference can clearly be seen in
fig.\ \ref{fig:BRplot}.
In fig.\ \ref{fig:BRplot} we plot the predicted value of
$\mathcal{B}(B_s \to \mu^+\mu^-)$ in the 2HDM as a function of 
$M_{H^+}$, for various values of $\tan\beta$.  For comparison we 
show the constraint on $M_{H^+}$ from $b \to s \gamma$ \cite{btosgamma},
obtained from the current 95\% confidence level experimental upper bound
of $\mathcal{B}(b \to s \gamma) < 4.5 \times 10^{-4}$ from the 
CLEO experiment.
For very large $\tan\beta$ and relatively light $H^+$, the 2HDM contribution
dominates and the branching fraction is significantly enhanced 
compared to its SM value.
As the 2HDM contribution becomes smaller due to increasing $M_{H^+}$
or decreasing $\tan\beta$, the branching fraction drops, eventually 
falling below the SM prediction due to the destructive interference.
If we ignore the kinematical factor of 
$(1 - 4m_{\ell}^2/m_{B_{d^{\prime}}}^2)$ in front of $C_S$ in \eq{eq:BR}
(which is a good approximation for $B_{d^{\prime}} \to \mu^+ \mu^-$ 
but not for $B_{d^{\prime}} \to \tau^+ \tau^-$)
then we can easily show that
the branching fraction in the 2HDM crosses the SM value when
\begin{equation}
	\frac{m^2_{B_s} \tan^2\beta}{8 M_W^2} \frac{\log r}{r-1}
	= Y(x_t),
\end{equation}
and reaches a minimum of half the SM value when
\begin{equation}
	\frac{m^2_{B_s} \tan^2\beta}{8 M_W^2} \frac{\log r}{r-1}
	= \frac{1}{2} Y(x_t).
\end{equation}
These correspond to $\tan^2 \beta \log r /(r-1) = 1790$ and 893, respectively.
As a numerical example, taking $\tan\beta = 60$ and $M_{H^+} = 175$ GeV 
(500 GeV),
we find $\mathcal{B}(B_s \to \mu^+ \mu^-) = 1.8 \times 10^{-8}$ 
($2.1 \times 10^{-9}$).

\begin{figure}[tbp]
	\begin{center}
	\resizebox{14cm}{!}
	{\rotatebox{270}
	{\includegraphics{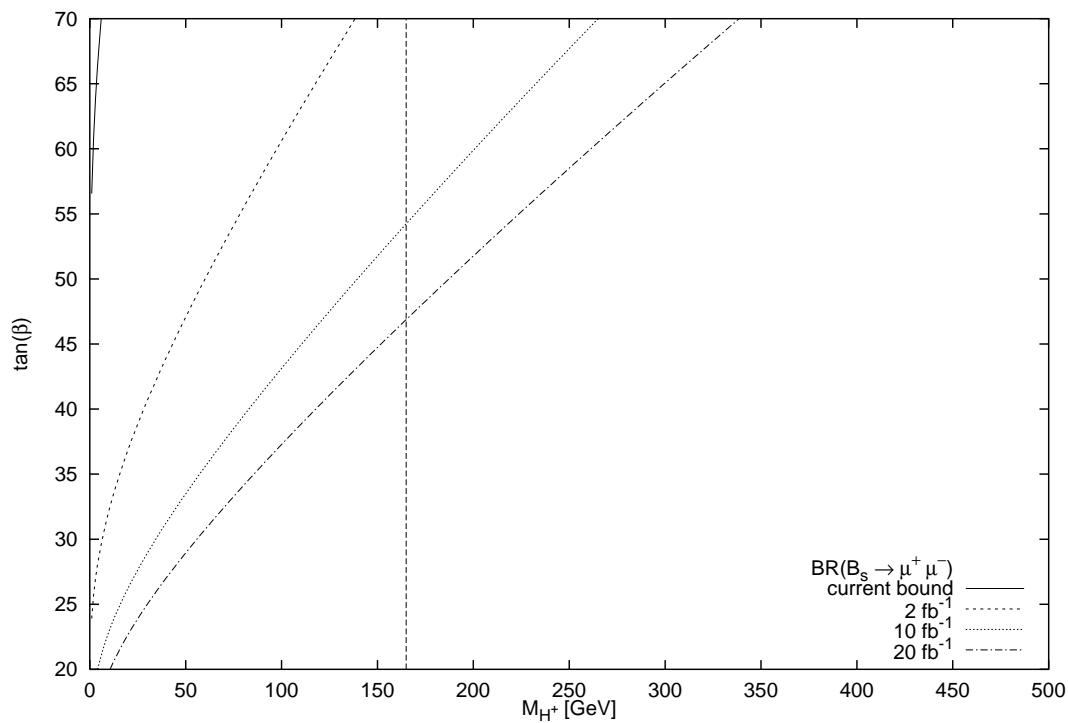}}}
	\end{center}
	\caption{Regions of $M_{H^+}$ and $\tan\beta$ parameter space
probed by measurements of $\mathcal{B}(B_s \to \mu^+\mu^-)$.
Contours are chosen as described in the text.
The vertical dashed line is the present lower bound on $M_{H^+}$
in the type-II 2HDM from $b \to s \gamma$ \cite{btosgamma}.}
	\label{fig:exclusionplot}
\end{figure}

In fig.\ \ref{fig:exclusionplot}
we plot the regions of $M_{H^+}$ and $\tan\beta$ parameter space 
that will be probed as the sensitivity to the decay $B_s \to \mu^+\mu^-$
improves at the Tevatron Run II.  
The contours shown (from left to right) were chosen as follows.
The current upper bound on $\mathcal{B}(B_s \to \mu^+ \mu^-)$ is 
$2.6 \times 10^{-6}$ from CDF \cite{cdf} with about 100 pb$^{-1}$ of integrated
luminosity.  This bound excludes a small region of parameter space 
with very high $\tan\beta$ and very light $H^+$, shown by the solid line
at the far left of fig.\ \ref{fig:exclusionplot}.  Such low $H^+$ masses
are already excluded by the constraint from $b \to s \gamma$ 
\cite{btosgamma}.  The rest of the contours in fig.\ \ref{fig:exclusionplot}
show the regions that we expect to be probed at the Tevatron Run II and 
extended Run II with various amounts of integrated 
luminosity, assuming
that the background for this process remains negligible.  In this case
the sensitivity to the branching fraction should scale with the luminosity.  
If there is background however, then the sensitivity will scale only
with the square root of the luminosity.
With 2 fb$^{-1}$ from each of the two detectors, the sensitivity 
should improve by a factor of 40
over the current sensitivity, to $6.5 \times 10^{-8}$, 
shown by the short dashes in fig.\ \ref{fig:exclusionplot}.  
For the values of $\tan\beta$ that
we consider, this sensitivity will still only probe values of $M_{H^+}$
already excluded by $b \to s \gamma$.  
We also show two contours for the expected sensitivity with 10 fb$^{-1}$
and 20 fb$^{-1}$ of integrated luminosity per detector
(dotted and dot-dashed lines in fig.\ \ref{fig:exclusionplot}, respectively).
These correspond to an extended Run II of the Tevatron.
With 10 fb$^{-1}$ we expect the sensitivity to reach a branching fraction
of $1.3 \times 10^{-8}$, allowing one to begin to probe $H^+$ masses above
the current bound from $b \to s \gamma$ for $\tan\beta > 54$.
With 20 fb$^{-1}$ we expect a reach of $6.5 \times 10^{-9}$, less than 
a factor of two above the predicted SM branching fraction.  This would
allow one to probe $H^+$ masses above the current bound from $b \to s \gamma$
for $\tan\beta > 47$.  In particular, for $\tan\beta = 60$, 
a non-observation of $B_s \to \mu^+ \mu^-$ at this sensitivity would
rule out $H^+$ lighter than 260 GeV.

Looking farther into the future, the experiments at the CERN LHC expect 
to observe the following numbers of signal (background) events for
$B_s \to \mu^+ \mu^-$ after 
three years of running at low luminosity, assuming the SM branching 
fraction \cite{LHCBreport}: ATLAS: 27 (93); CMS: 21 (3);
and LHCb: 33 (10).  Since the suppression of this branching fraction 
in the 2HDM is at most a factor of two, the LHC experiments will be
able to observe this decay for any configuration of the 2HDM at 
large $\tan\beta$.
For e.g.\ $\tan\beta = 60$ and $M_{H^+} < 285$ GeV, the branching fraction 
in the 2HDM is enhanced by 30\% or more compared to the SM.
Similarly, for $\tan\beta = 60$ and 375 GeV $< M_{H^+} <$ 1 TeV, 
the branching fraction
is suppressed by 30\% or more compared to the SM\footnote{We 
do not consider
charged Higgs masses larger than 1 TeV for naturalness reasons.}.  In 
these regions, we expect the LHC to be able to distinguish the 2HDM from
the SM.  In the region of large $M_{H^+}$, the dependence on $M_{H^+}$
is very weak; hence the LHC measurement will give powerful constraints on 
$\tan \beta$ in the large $\tan \beta$ region.  

We have made no attempt to simulate the experimental background for
this decay in order to obtain an accurate estimate of the reach of 
the Tevatron Run II.  Neither have we taken into account the 
theoretical uncertainty.  We expect the largest theoretical 
uncertainty to come from uncertainties in the input parameters,
primarily the $B$ meson decay constants and CKM matrix elements 
in \eq{eq:numBRs}.
These uncertainties will be reduced as the $B$ physics experiments progress
and lattice calculations improve.
Also QCD corrections to the 2HDM contribution will arise at NLO and require
the calculation of two-loop diagrams.  In the SM, the NLO corrections
increase the decay amplitude by about 3\%, and therefore increase the 
branching fraction by about 6\%.  We expect the NLO corrections
to the 2HDM contribution to be of the same order, in which case our 
conclusions are not significantly modified.

In order to evaluate the usefulness of $B_s \to \mu^+ \mu^-$ as a 
probe of the 2HDM, we must compare it to other measurements that 
constrain the 2HDM in the large $\tan\beta$ regime.
As the statistics of $B$ physics experiments improve, the measurement
of $b \to s \gamma$ will improve as well.  If $B_s \to \mu^+ \mu^-$ 
is to be a useful probe of the 2HDM, it must be sensitive to a range 
of parameter space not already explored by $b \to s \gamma$ at each
integrated luminosity.  Fortunately,
$B_s \to \mu^+ \mu^-$ is complementary to $b \to s \gamma$ because of
the different dependence on $\tan\beta$.  While the 2HDM contributions
to $b \to s \gamma$ are independent of $\tan\beta$ for $\tan\beta$ larger
than a few, the 2HDM contributions to $B_s \to \mu^+ \mu^-$ depend 
sensitively on $\tan\beta$.  This makes $B_s \to \mu^+ \mu^-$ an 
especially sensitive probe of the 2HDM in the large $\tan\beta$ regime,
while $b \to s \gamma$ will remain more sensitive for moderate and small
$\tan\beta$.
Finally, a fit to the $Z$ decay data in the 2HDM \cite{olegetal} 
puts weak constraints on the $H^+$ mass for very large $\tan\beta$:
$M_{H^+} > 40$ GeV at 95\% confidence level for $\tan\beta = 100$.
The fit gives no constraint for $\tan\beta < 94$.


\section{Conclusions}
\label{sec:conclusions}
In this paper we have analyzed the decays $B_{d^{\prime}} \to \ell^+ \ell^-$
in the type-II 2HDM with large $\tan\beta$.  Although these decays
have been studied in a 2HDM before, the previous analyses omitted the 
box diagram involving $W$ and $H^+$, which is the dominant
contribution at large $\tan\beta$ in the 't~Hooft-Feynman gauge and is
needed for gauge independence.  We showed that when all the contributions
are properly included in the large $\tan\beta$ limit, the resulting 
expressions for the branching fractions are quite simple and depend 
only on $\tan\beta$ and the charged Higgs mass.  
These 2HDM contributions can enhance or suppress the branching fractions 
by a significant amount compared to their SM values, providing
tantalizing search possibilities with the potential to probe
large parts of the large $\tan\beta$ parameter space of the 2HDM.
We have focused in our numerical analysis on $B_s \to \mu^+ \mu^-$,
for which the the experimental sensitivity is best.  We find that
for reasonable values of $\tan\beta$ up to 60 and charged Higgs masses
above the lower bound set by $b \to s \gamma$, 
$\mathcal{B}(B_s \to \mu^+ \mu^-)$ can be increased by 
up to a factor of five 
above its SM expectation or suppressed by up to a factor of
two, depending on the charged Higgs mass.  Although very high statistics
will be needed to observe this decay, it promises to be an 
experimentally and theoretically clean probe of new Higgs physics.

\vskip1cm

{\Large \bf Acknowledgments}
\vskip0.5cm
We are grateful to Jan Kalinowski and Witold Skiba for helpful discussions,
and to Jonathan Lewis for discussions on the reach of the Tevatron in Run II.
We would also like to thank the conveners of the B-Physics at Run II 
Workshop at Fermilab for facilitating useful interaction among theorists
and experimentalists.
Finally we owe a debt of gratitude to Piotr Chankowski and
\L ucja S\l awianowska for pointing out 
an error in the relative sign between the SM- and 2HDM-induced
contributions to the decay rate
in an earlier version of this manuscript,
and for confirming our result for the 2HDM Wilson coefficients.

\appendix

\section{Trilinear Higgs couplings}
\label{app:trilinear}
The trilinear $H^+H^-H$ couplings ($H=h^0,H^0$) for the non-supersymmetric 
2HDM were first presented in \cite{sk}.  These couplings depend strongly on
the model considered.  For the most general CP-conserving 2HDM with 
natural flavor conservation, the $H^+H^-H$ couplings are given by 
$igQ_H/M_W$, where
\begin{eqnarray}
	Q_{h^0} &=& -v_{SM}^2 \lt[ 
	\sin(\beta - \alpha)(2\lambda_3 + \lambda_4)
	- \sin\beta\cos\beta\cos(\alpha + \beta) \lambda_5 \rt.
	\nonumber \\
	&& \qquad \lt. + 2 \sin\beta\cos\beta 
	\lt(-\sin\alpha\sin\beta \lambda_1
	+ \cos\alpha\cos\beta \lambda_2 \rt) \rt]
	\nonumber \\
	Q_{H^0} &=& -v_{SM}^2 \lt[
	\cos(\beta - \alpha)(2\lambda_3 + \lambda_4)
	- \sin\beta \cos\beta \sin(\alpha + \beta) \lambda_5 \rt.
	\nonumber \\
	&& \qquad \lt. + 2 \sin\beta\cos\beta 
	\lt( \cos\alpha\sin\beta \lambda_1
	+ \sin\alpha\cos\beta \lambda_2 \rt) \rt].
	\label{Qh}
\end{eqnarray}
The $H^+H^-A^0$ coupling is zero.
Here $v_{\rm SM} = 174$ GeV is the SM Higgs vev, and the 
$\lambda_i$ are the scalar quartic couplings of the Higgs potential 
given in \cite{HHG}.  To write these couplings in terms of Higgs masses
and mixing angles, one must make an assumption to eliminate one of the
independent $\lambda_i$.  In \cite{sk}, formulae are presented for the 
two cases $\lambda_1 = \lambda_2$ and $\lambda_5 = \lambda_6$.  The
formulae in \eq{Qh} agree with \cite{sk} in these two cases.

At large $\tan\beta$, \eq{Qh} reduces to
\begin{eqnarray}
	Q_{h^0} &\simeq& -v^2_{\rm SM} \cos\alpha
	(2\lambda_3 + \lambda_4) \, [1 + \mathcal{O}(\cot\beta)] \nonumber \\
	Q_{H^0} &\simeq& -v^2_{\rm SM} \sin\alpha
	(2\lambda_3 + \lambda_4) \, [1 + \mathcal{O}(\cot\beta)].
	\label{Qexpansions}
\end{eqnarray}
These couplings are not enhanced at large $\tan\beta$.

Considering instead the case $\lambda_1 = \lambda_2$ and writing the
trilinear couplings in terms of Higgs masses and mixing angles, we 
find at large $\tan\beta$,
\begin{eqnarray}
	Q_{h^0} &\simeq& - \nicefrac{1}{2} M^2_{H^0} 
	\tan\beta \cos^2\alpha \sin\alpha \, [1 + \mathcal{O}(\cot\beta)] 
	\nonumber \\
	Q_{H^0} &\simeq& \nicefrac{1}{2} M^2_{h^0} 
	\tan\beta \cos\alpha \sin^2\alpha \, [1 + \mathcal{O}(\cot\beta)].
	\label{Qmh}
\end{eqnarray}
Naively, one would conclude that these couplings are enhanced at large
$\tan\beta$.  This is incorrect because the angle $\alpha$ depends
on $\tan\beta$.  At large $\tan\beta$ we have
\begin{equation}
	\tan 2\alpha = \frac{2(4\lambda_3 + \lambda_5)}
	{\lambda_5 - 4(\lambda_2 + \lambda_3)} \cot\beta \,
	[1 + \mathcal{O}(\cot^2\beta)].
\end{equation}
Thus for generic values of the $\lambda_i$, $\sin\alpha \sim \cot\beta$,
and the $\tan\beta$ enhancement in \eq{Qmh} is cancelled.

\end{document}